\documentclass[12pt]{article}
\usepackage{epsfig,enumerate,color,fullpage}
\def\myendproof{{\ \vbox{\hrule\hbox{%
   \vrule height1.3ex\hskip0.8ex\vrule}\hrule }}\par}
\newtheorem{theorem}{Theorem}[section]
\newtheorem{lemma}[theorem]{Lemma}
\newtheorem{corollary}[theorem]{Corollary}
\newenvironment{proof}{{\it Proof. }}{\myendproof}

\begin{document}

\title{\Large\bf On the Ramsey Numbers for Bipartite Multigraphs\footnote{Research supported in part by NSC grants NSC-91-2815-C-002-092-E.}}

\author{Ming-Yang Chen\thanks{Department of Electrical Engineering, National Taiwan University,
Taipei 106, Taiwan, Republic of China. Email: roc3.chen@msa.hinet.net.}
\and
Hsueh-I Lu\thanks{Corresponding author: Institute of Information Science, Academia Sinica,
Taipei 115, Taiwan, Republic of China. Email: hil@iis.sinica.edu.tw. URL: www.iis.sinica.edu.tw/\~{ }hil/}
\and
Hsu-Chun Yen\thanks{Department of Electrical Engineering, National Taiwan University
Taipei 106, Taiwan, Republic of China. Email: yen@cc.ee.ntu.edu.tw. URL: www.ee.ntu.edu.tw/\~{ }yen/}
}

\date{May 12, 2003}
\maketitle

\begin{abstract}
A coloring of a complete bipartite graph is shuffle-preserved if it is
the case that assigning a color $c$ to edges $(u, v)$ and $(u', v')$
enforces the same color assignment for edges $(u, v')$ and
$(u',v)$. (In words, the induced subgraph with respect to color $c$ is
complete.)  In this paper, we investigate a variant of the Ramsey
problem for the class of complete bipartite multigraphs. (By a
multigraph we mean a graph in which multiple edges, but no loops, are
allowed.) Unlike the conventional m-coloring scheme in Ramsey theory
which imposes a constraint (i.e., $m$) on the total number of colors
allowed in a graph, we introduce a relaxed version called m-local
coloring which only requires that, for every vertex $v$, the number of
colors associated with $v$'s incident edges is bounded by $m$. Note
that the number of colors found in a graph under $m$-local coloring
may exceed m.  We prove that given any $n \times n$ complete bipartite
multigraph $G$, every shuffle-preserved $m$-local coloring displays a
monochromatic copy of $K_{p,p}$ provided that $2(p-1)(m-1) < n$.
Moreover, the above bound is tight when (i) $m=2$, or (ii) $n=2^k$ and
$m=3\cdot 2^{k-2}$ for every integer $k\geq 2$. As for the lower bound
of $p$, we show that the existence of a monochromatic $K_{p,p}$ is not
guaranteed if $p> \left\lceil \frac{n}{m}\right\rceil$. Finally, we
give a generalization for $k$-partite graphs and a method applicable
to general graphs.  Many conclusions found in $m$-local coloring can
be inferred to similar results of $m$-coloring.
\end{abstract}

\section{Introduction}

{\em Ramsey theory}, originated in a seminal paper by
Ramsey~\cite{R30} in 1930, has emerged as a fast growing and
fascinating research topic in mathematics and theoretical computer
science in recent years.  Ramsey theory deals with the investigation
of the conditions under which a sufficiently large complete graph
always includes a certain substructure~\cite{GRS80}.  Among various
graphs for which Ramsey-type problems have been investigated,
bipartite graphs constitute a class for which several deep results
have been obtained for a variety of Ramsey numbers (see,
e.g.,~\cite{AFM00,CG75,ER93,CC99}).  For example, in \cite{ER93},
Erd\H{o}s and Rousseau proved that in every 2-coloring of $K_{n,n}$,
there is a monochromatic copy of $K_{p,p}$ if
\begin{displaymath}
  n {n/2\choose p} > (p-1){n \choose p}.
\end{displaymath}
In a more recent article~\cite{CC99}, Carnielli and Carmelo showed
that in any 2-coloring of a bipartite complete graph $K_{n,n}$, one
can always find a monochromatic subgraph isomorphic to $K_{p,q}$ if $n
\geq 2^{p}(q-1)+2^{p-1}-1$. As in asymptotic
versions~\cite{CaroR01,LiRCZ01}, Caro and Rousseau achieved that there
are constants $c_1$ and $c_2$ such that
$$c_1\left( \frac{p}{\log p}\right)^{(q+1)/2} < b(p,q) < c_2\left(
\frac{p}{\log p}\right)^{q},$$ holds as $p$ tends to infinity, where
$b(p,q)$ is the smallest integer $n$ such that every 2-coloring (say
red and green) of the edges of $K_{n,n}$ contains either a red
$K_{p,p}$ or a green $K_{q,q}$.

\begin{figure}[ht]
\label{example-1}
\centerline{\input{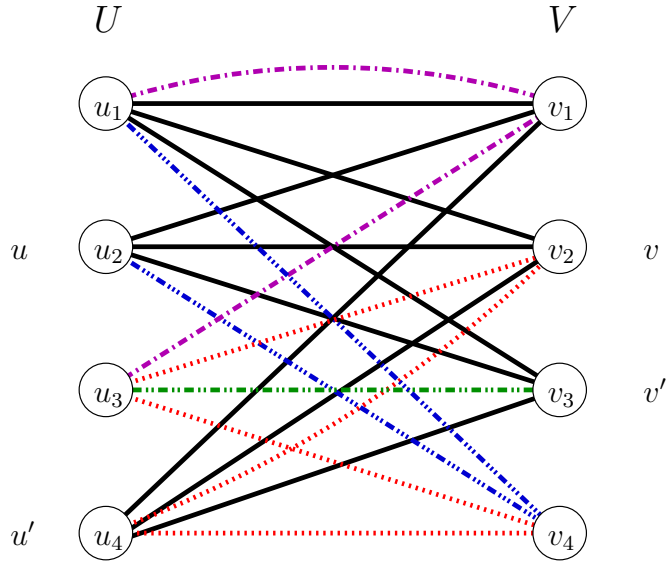}}
\caption{A monochromatic $K_{3,3}$ in a coloring of a $4 \times 4$
bipartite multigraph.  Notice that here $m$ is bounded by 3, although
the total number of colors used in the graph is 5.}
\end{figure}

Some recent research focuses turn to seek other types of Ramsey
theory which give tighter bounds or stronger relations with other
well-known combinatorial numbers, but usually require additional
constraints on the graph.  For example, Alon, Erd\H{o}s, Gunderson,
and Molloy~\cite{AlonEGM02} proved that the corresponding largest
integer $m$ is asymptotically equal to the Tur\`{a}n number $t(n,
\lfloor {n \choose 2}/k\rfloor)$ if the smallest Ramsey number $n$
satisfies that, for any $k$-coloring of complete graph $K_{n}$, there
exists a copy of $K_{m}$ whose edges receive at most $k-1$ colors.
Similarly, we address how to find a monochromatic copy of biclique in
a bipartite multigraph with a {\em shuffle-preserved} coloring, i.e., a
coloring $\eta$ such that for all $u, u' \in U$ and $v, v' \in V$, the
condition $(u, v)_{c}\wedge (u', v')_{c}$ implies the condition $(u,
v')_{c} \wedge (u', v)_{c}$, where $(u, v)_{c}$ denotes the existence
of an edge colored $c$ between vertices $u$ and $v$.  (See Figure
\ref{example-1}.)  In the present paper, we prove that given any $n
\times n$ complete bipartite multigraph $G$, every shuffle-preserved
$m$-local coloring displays a monochromatic copy of $K_{p,p}$ provided
that $2(p-1)(m-1) < n$.  Moreover, the above bound is tight when (i)
$m=2$, or (ii) $n=2^k$ and $m=3\cdot 2^{k-2}$ for every integer $k\geq
2$. As for the lower bound of $p$, we show that the existence of a
monochromatic $K_{p,p}$ is not guaranteed if $p> \left\lceil
\frac{n}{m}\right\rceil$. Finally, we give a generalization for
$k$-partite graphs and a method applicable to general graphs.

It is worthy of pointing out that the constraint of
shuffle-preserving, to a certain extent, resembles the requirement
found in the notion of {\em induced Ramsey numbers} (see, e.g.,
\cite{KPR98}) which deals with conditions under which the existence of
a monochromatic induced subgraph is guaranteed.  Note that what
shuffle-preserved means is that for every color $c$, the subgraph
induced by $c$ is complete.  (A subgraph $(V',E')$ of $(V, E)$ (where
$V' \subseteq V$ and $E' \subseteq E$) is said to be {\em induced} if
$E'=\{(u,v) : u, v \in V', (u,v) \in E\}$.) As we shall see later, the
underlying coloring being shuffle-preserved plays a critical role in
the existence of a much smaller Ramsey number, in comparison with
those reported in the literature for bipartite graphs. For more about
Ramsey numbers for a variety of graphs, the interested reader is
referred to~\cite{Ra02}.

The rest of the paper is organized as follows.
Section~\ref{section:ramsey} presents the key theorem of this
paper, whose tightness is addressed in
Section~\ref{section:tightness}.
Section~\ref{section:generalization} shows our results for
$k$-partite graphs and a generalization to  general graphs.
Section~\ref{section:conclusion} concludes the paper with some
open questions.

\section{Ramsey theory for bipartite multigraphs}
\label{section:ramsey}
An $n \times n$ bipartite multigraph $G$ is said to be {\em complete}
if every two vertices of different bipartitions are adjacent.  Given a
vertex $u$, we write $C(u)$ to denote the set $\{c \in C : \mbox{an
incident edge of $u$ is colored $c$}\}$.  For an $n \times n$
bipartite multigraph $G=(V, E)$ and a color set $C$, a coloring $\eta$
is said to be an {\em $m$-local coloring} if for every vertex $u$,
$|C(u)| \leq m$ (i.e., the number of colors assigned to $u$'s incident
edges is at most $m$).  Notice that for any $m$-local coloring, the
total number of colors used in the entire graph may exceed $m$.  Now
the following main theorem gives Ramsey theory for bipartite
multigraphs.

\begin{theorem}
\label{thm1} Let $G$ be an $n\times n$ complete bipartite
multigraph.  If $p$ and $m$ are positive integers such that
$2(p-1)(m-1)<n$, then any shuffle-preserved $m$-local coloring has
a monochromatic copy of $K_{p,p}$.
\end{theorem}

\begin{proof}
Suppose that the vertex set of $G$ is partitioned as $U \cup V$.  Let $\eta$ be
an arbitrary shuffle-preserved $m$-local coloring on $G$ using color
set $C$.  Without loss of generality, we may assume that all the
multiple edges between each pair of vertices have distinct colors
under $\eta$.  We define $U(c)$ (respectively, $V(c)$) to be the set
of vertices in $U$ (respectively, $V$) each of which has at least one
incident edge colored $c$.  To prove our theorem, it suffices to show
that there exists a $c\in C$ such that $\left| {U(c)}\right| \geq p$
and $\left| {V(c)}\right| \geq p$.  (In this case, the existence of a
monochromatic $K_{p,p}$ follows immediately from $\eta$ being
shuffle-preserved.)

We prove the theorem by contradiction.  Suppose, on the contrary, that
there were no $c\in C$ satisfying $\left| {U(c)}\right| \geq p$ and
$\left| {V(c)}\right| \geq p$.  For convenience, we write the ordered
triple $(u,v,c)$ to represent the relation $c\in C(u)\cap C(v)$, and
let $T$ be the set of all such triples.  By $K_{n,n}\subseteq G$, it
is reasonably easy to observe
\begin{equation}
\label{eq1}
\left| T\right| =\sum_{u\in U}\sum_{v\in V}\left| C(u) \cap
C(v) \right| \geq \left| U\right| \cdot \left| V\right| =n^{2}.
\end{equation}
Furthermore, by changing order in
double summation, the following also hold.
\begin{equation}
\label{eq2}
\sum_{c\in C}\left| U(c) \right|  = \sum_{u\in U}\left| C(u) \right|
  \leq m\left| U\right|
\end{equation}

\begin{equation}
\label{eq3}
\sum_{c\in C}\left| V(c) \right| = \sum_{v\in V}\left| C(v) \right|
\displaystyle
\leq m\left| V\right|.
\end{equation}
Let
\begin{eqnarray*}
C_{1}&=&\left\{c\in C : \left| U(c) \right| \geq p\right\};\\
C_{2}&=&\left\{c\in C : \left| U(c) \right| <p\right\}.
\end{eqnarray*}
For a vertex $u$ and a color $c$, we write
$u\stackrel{c}{\rightarrow}$ to represent the set of $u$'s adjacent
vertices each of which is connected to $u$ through some edge of color
$c$.  For every $u \in U$, by the Pigeonhole Principle there must be a
color $c \in C$ such that $\left|u\stackrel{c}{\rightarrow}\right|
\geq \lceil n/m \rceil$ (i.e., at least $\lceil n/m \rceil$ incident
edges of $u$ are colored $c$).  Note that $p-1<\frac{n}{2(m-1)}$
implies
\begin{displaymath}
\left\lceil\frac{n}{2(m-1)}\right\rceil\geq p.
\end{displaymath}
By $m \geq 2$ we have
\begin{displaymath}
\left|u\stackrel{c}{\rightarrow}\right| \geq \left\lceil \frac{n}{m}\right\rceil
\geq p.
\end{displaymath}
Hence, $\left| V(c) \right| \geq \left|u\stackrel{c}{\rightarrow}\right| \geq p$
(at least $p$ of vertex $u$'s neighbors
are in $V(c)$).  Due to our assumption that no $c \in C$ satisfies
$\left| {U(c)}\right| \geq p$ and $\left| {V(c)}\right| \geq p$
simultaneously, $c$ must be in $C_{2}$.  For every vertex $u \in U$,
there exists a $c \in C(u)$ such that $c \in C_{2}$.  Hence,
\begin{displaymath}
\sum_{c \in C_{2}} \left| U(c) \right| \geq \left| U \right|
\end{displaymath}
and from Equations~(\ref{eq2}) and~(\ref{eq3}) we have
\begin{displaymath}
\sum_{c \in C_{1}} \left| U(c) \right| \leq (m-1) \left| U
\right|.
\end{displaymath}
Similarly for each $v \in V$, there must be a color $c$ such that
$\left|v\stackrel{c}{\rightarrow}\right| \geq p$ (i.e., at least $p$ vertices in
$U$ are adjacent to $v$ through edges colored $c$).  Thus $c$ will be
in $C_{1}$ and we have
\begin{eqnarray*}
\sum_{c \in C_{1}} \left| V(c) \right| &\geq& \left| V \right|;\\
\sum_{c \in C_{2}} \left| V(c) \right| &\leq& (m-1) \left| V \right|.
\end{eqnarray*}
Therefore, using the above inequalities, we find
\begin{eqnarray*}
\left| T \right|
&=&\sum_{c\in C}\left| U(c) \right| \cdot \left| V(c) \right|\\
&=&\sum_{c\in C_{1}}\left| U(c) \right| \cdot \left| V(c) \right|
  + \sum_{c\in C_{2}}\left| U(c) \right| \cdot \left| V(c) \right|\\
&\leq&(p-1) \sum_{c\in C_{1}}\left| U(c) \right|
  +(p-1)\sum_{c\in C_{2}} \left| V(c) \right| \\
&\leq& (p-1)(m-1) \left| U\right| +(p-1)(m-1) \left| V\right| \\
&=& 2(p-1)(m-1)n\\
&<&n^{2},
\end{eqnarray*}
which contradicts Equation~(\ref{eq1}). Our theorem follows.
\end{proof}

Since every {\em m-coloring} (i.e., coloring a graph with at most
$m$ distinct colors) is clearly an $m$-local coloring, the
following is straightforward.

\begin{corollary}
\label{coro-1}
In every shuffle-preserved $m$-coloring of an $n \times n$ complete
bipartite multigraph $G$, there is a monochromatic copy of $K_{p,p}$
if
\begin{displaymath}
2(p-1)(m-1)< n.
\end{displaymath}
\end{corollary}
For $m=2$, Corollary~\ref{coro-1} suggests a sufficient condition of
$n > 2(p-1)$ for the existence of a monochromatic $K_{p,p}$.  The
interested reader should contrast this result with a much larger bound
(i.e., $n > 2^{p}p$) in~\cite{ER93} in which the
shuffle-preserved constraint is lifted.

\section{Necessary and sufficient conditions}
\label{section:tightness} In this section we provide necessary and
sufficient conditions for the existence of a monochromatic copy of
$K_{p,p}$ for some special cases.

\begin{lemma}
\label{lemma:thm2}
Let $G$ be a complete $n\times n$ bipartite graph without multiple
edges.
\begin{enumerate}
\item
If $p>\left\lceil\frac{n}{m}\right\rceil$, then there exists a
shuffle-preserved $m$-coloring $($and thus a shuffle-preserved
$m$-local coloring$)$ of $G$ such that $G$ does not contain any
monochromatic copy of $K_{p,p}$.
\item
If $n=2^{k}$, $m=3\times 2^{k-2}$, and $p=2$ for some $k \geq 2$, then
there exists a shuffle-preserved $m$-local coloring of $G$ such that
$G$ does not contain any monochromatic copy of $K_{p,p}$.
\end{enumerate}
\end{lemma}

\begin{proof}
Let $u_i$ (respectively, $v_i$) be the $i$-th node of $U$
(respectively, $V$).

\begin{figure}
\centerline{\input{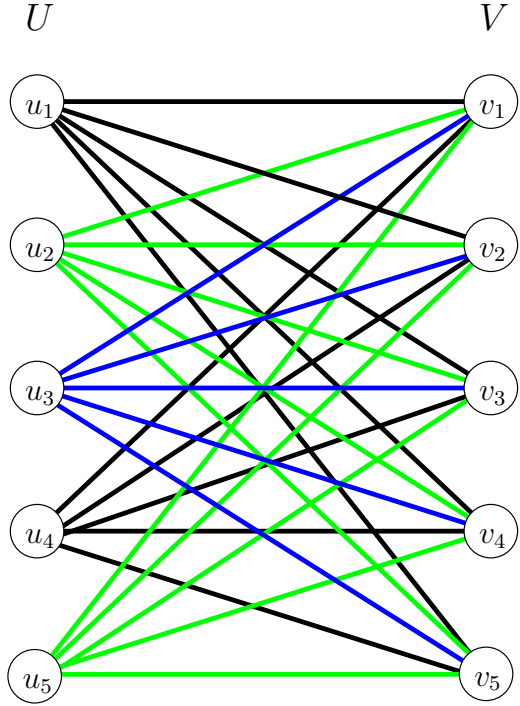}}
\caption{An example which does not contain any monochromatic copy of
$K_{3,3}$ for $(n,m,p)=(5,3,3)$ in Lemma~\ref{lemma:thm2}(1).}
\label{fig:fig2}
\end{figure}

Statement 1.
We color the edge between $u_i$ and $v_j$ by
color $(i\bmod m)$.  One can easily verify that such a coloring is
indeed a shuffle-preserved $m$-coloring. By $p > \left\lceil
\frac{n}{m}\right\rceil$, each $v_i$ has at most $p-1$ incident
edges with the same color. Therefore, $G$ contains no monochromatic
copy of $K_{p,p}$. (See Figure~\ref{fig:fig2}.)

\begin{figure}
\centerline{\input{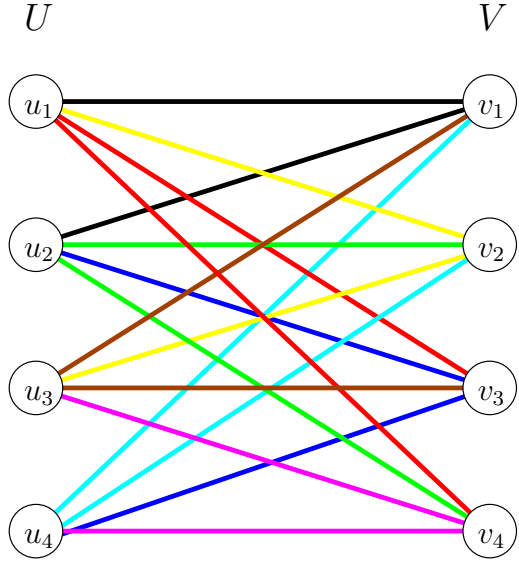}}
\caption{The edge coloring of the matrix $M_2$ in
Lemma~\ref{lemma:thm2}(2) shows that no monochromatic subgraph of
$K_{2,2}$ in the case $(n,m)=(4,3)$.}
\label{fig:fig3}
\end{figure}

Statement 2.  For each $\ell=2,3,\ldots, k$, let $M_\ell$ denote the
$2^{\ell} \times 2^{\ell}$ color matrix whose the $(i,j)$-entry
specifies the color of the edge between $u_i$ and $v_j$.  We
construct $M_k$ recursively by letting
\begin{displaymath}
M_2 =
\begin{tabular}{|c|c|c|c|} \hline
1 & 5 & 2 & 2 \\ \hline
1 & 4 & 3 & 4 \\ \hline
8 & 5 & 8 & 7 \\ \hline
6 & 6 & 3 & 7 \\ \hline
\end{tabular}
\end{displaymath}
(see Figure~\ref{fig:fig3}) and
\begin{displaymath}
M_{\ell+1}=
\begin{tabular}{|c|c|} \hline
$M_{\ell}$ & $\mu_\ell+ M_{\ell}$ \\ \hline
$2\mu_\ell+ M_{\ell}$ & $3\mu_\ell+M_{\ell}$ \\ \hline
\end{tabular}
\end{displaymath}
for each $\ell=3,4,\ldots,k$, where $t\mu_\ell + M_{\ell}$ denotes the
matrix obtained by adding $t$ times of the maximum of $M_\ell$ to each
entry of $M_{\ell}$.  Clearly, $M_{2}$ is
 3-local coloring since each row (and column) consists of
three colors.  The lack of a monochromatic $K_{2,2}$ in
 $M_{2}$ is  also straightforward.
Likewise, it is reasonably easy to  verify that the constructed $M_k$ gives
a shuffle-preserved $3 \times 2^{k-2}$-local coloring of $G$ such that no
monochromatic copy of $K_{p,p}$ can be found in $G$.
\end{proof}

Combining Theorem~\ref{thm1} and Lemma~\ref{lemma:thm2}, we have
the following results.

\begin{theorem}
\label{thm5}
Every shuffle-preserved 2-local coloring of an $n \times n$ complete
bipartite graph $G$ gives a monochromatic copy of $K_{p,p}$ if and
only if
\begin{displaymath}
2(p-1)<n.
\end{displaymath}
\end{theorem}

\begin{proof}
Since $(p-1) \geq \frac{n}{2}$ and $p>\left\lceil
\frac{n}{2}\right\rceil$ are equivalent, the theorem follows from
Theorem~\ref{thm1} and Lemma~\ref{lemma:thm2}(1) with
$m=2$.\footnote[2]{Observe that Theorem~\ref{thm5} also holds even if
changing 2{\em-local coloring} to 2{\em-coloring} due to the nature
of Lemma~\ref{lemma:thm2}(1).}
\end{proof}

\begin{theorem}
\label{thm6}
Let $n=2^{k}$ and $m=3\times 2^{k-2}$ for some integer $k \geq 2$.
Any shuffle-preserved $m$-local coloring of an $n\times n$ complete
bipartite graph gives a monochromatic copy of $K_{p,p}$ if and only if
\begin{displaymath}
2(p-1)(m-1)< n.
\end{displaymath}
\end{theorem}

\begin{proof}
By Theorem~\ref{thm1} and Lemma~\ref{lemma:thm2}(1), the theorem holds when
$2(p-1)(m-1)<n$ or $p>\left\lceil\frac{n}{m}\right\rceil$. When
$2(p-1)(m-1)\geq n$ and $p\leq \left\lceil\frac{n}{m}\right\rceil$, by
$n=2^k$, $m=3\times 2^{k-2}$, and $k\geq 2$, one can obtain
$p=2$. Thus, the theorem follows from Lemma~\ref{lemma:thm2}(2).
\end{proof}

\section{Generalization}
\label{section:generalization}

In this section, we give a generalization of the results in
Section \ref{section:ramsey} for  $k$-partite graphs as well as
for  general graphs. For  complete $k$-partite graphs, the
following results generalize Corollary~\ref{coro-1} and
Lemma~\ref{lemma:thm2}(1).
\begin{corollary}
\label{coro-2}
In every shuffle-preserved $2$-coloring of an $n \times
n\times\cdots\times n$ complete $k$-partite multigraph, there is a
monochromatic copy of a complete $p \times p\times\cdots\times p$
$k$-partite graph $G$ if
\begin{displaymath}
2(p-1)< n.
\end{displaymath}
\end{corollary}
\begin{proof}
Let the $k$-partitions of the vertex set be
\begin{eqnarray*}
S_{1} &=& \{v_{1,1}, v_{1,2}, \ldots, v_{1,n}\};\\
S_{2} &=& \{v_{2,1}, v_{2,2}, \ldots, v_{2,n}\};\\
      &\vdots&\\
S_{k} &=& \{v_{k,1}, v_{k,2}, \ldots, v_{k,n}\}.
\end{eqnarray*}
The case $k=2$ follows from Corollary~\ref{coro-1}.  When $k=3$,
we can find one monochromatic $K_{p, p}$ in each bipartite
subgraph of $G$ induced by $S_i$ and $S_j$ with $i\ne j$.  Since
the graph is two-colored, from the Pigeonhole Principle at least
two monochromatic copies of $K_{p, p}$ have the same color, say
$c$.  This tells us that for every set $S_{i}$, there are at least
$p$ vertices, each of which has some incident edge colored $c$,
and thus from the coloring being shuffle-preserved, we have a
monochromatic $p \times p$ tripartite graph.

Suppose now $k > 3$. Let $G'_i$ denote the $p \times p\times
\cdots\times p$ $(k-1)$-partite subgraph of $G$ induced by all but the
nodes in $S_i$.  By the inductive hypothesis, there are $k$
monochromatic $p \times p\times\cdots\times p$ $(k-1)$-partite graphs
$G'_{1}, G'_{2}, \ldots , G'_{k}$. By the Pigeonhole Principle, at
least two of them have the same color.  By the same argument, a
monochromatic $p \times p\times\cdots\times p$ $k$-partite graph is
obtained.
\end{proof}

\begin{corollary}
If $p>\left\lceil\frac{n}{m}\right\rceil$, then there exists an
$m$-colored complete $n\times n\times \cdots \times n$ $k$-partite
multigraph $G$ that does not contain any monochromatic complete
$p\times p\times\cdots\times p$ $k$-partite subgraph.
\end{corollary}

\begin{proof}
The proof is a natural generalization of that for
Lemma~\ref{lemma:thm2}(1).  Let $S_1,S_2,\ldots,S_k$ be the
$k$-partitions of the vertex set of $G$.  The edges of $G$ are
constructed and colored as follows.  Let the subgraph of $G$ induced
by $S_1$ and $S_j$ for any $j$ with $2\leq j\leq k$ contain no
multiple edges. We color all the incident edges of the $i$-th node of
$S_1$ by color $(i\bmod m)$.  For any indices $i$ and $j$ with $2\leq
i<j\leq k$, let the subgraph of $G$ induced by $S_i$ and $S_j$ be $m$
superimposed copies of $K_{n,n}$, each with a distinct color. It is
not difficult to verify that the resulting coloring is
shuffle-preserved. By $p>\left\lceil\frac{m}{n}\right\rceil$, any node
in $S_2\cup S_3\cup \cdots \cup S_k$ is adjacent to at most $p-1$
nodes in $S_1$ through edges with the same color. Therefore, $G$ does
not contain any complete $p\times p\times \cdots\times p$ $k$-partite
subgraph.
\end{proof}

Hence, we get the following theorem immediately.

\begin{theorem}
\label{thm7}

In every shuffle-preserved $2$-coloring of an $n \times
n\times\cdots\times n$ complete $k$-partite multigraph, there is a
monochromatic copy of a complete $p \times p\times\cdots\times p$
$k$-partite graph $G$ if and only if
\begin{displaymath}
2(p-1)< n.
\end{displaymath}
\end{theorem}

Now we consider general graphs. Clearly, we have to generalize the
definition of shuffle-preserved coloring as follows: a coloring $\eta$
is {\em shuffle-preserved} if the induced subgraph with respect to
every color $c$ is complete.  Recall that the classical Ramsey theory
focuses on the situation that the particular subgraph is required to
be monochromatic.  In the next theorem we give a sufficient condition
for a more general situation that the shuffle-preserved coloring
guarantees the existence of a complete subgraph whose edges receive at
most some number of colors.  (See~\cite{AlonEGM02} for a similar
theorem on complete graphs.)

\begin{theorem}
\label{thm8}

Let $G$ be a graph with $n$ vertices and shuffle-preserved $m$-coloring.
Suppose that $d_{i}$ represents the number of vertices, each
of which has exactly $i$ different colors among its incident
edges. Then $G$ has a $t$-superimposed copy of $K_{p}$ if

$$ p \leq \left\lceil \frac{\sum _{i=t}^{m}d_{i}{i \choose t}}{{m \choose t}} \right\rceil.$$
\end{theorem}

\begin{proof}
Let $\mathcal{P}_{t}(S)$ be the set of all $t$-element subsets of
$S$ and $V_{i}$ consist of the vertices of the complete subgraph
induced by the $i$-th color. Then by two-way counting we have
$\sum_{\{i_{1},i_{2},\ldots,i_{t}\} \in \mathcal{P}_{t}(\{1,2,\ldots, m\})} |V_{i_{1}}\cap V_{i_{2}}\cap \cdots \cap V_{i_{t}}| = \sum _{i=t}^{m}d_{i}{i \choose t}$
since both sides give the number of all vertices in any $t$-element combination
among $\{V_{1}, V_{2}, \ldots, V_{m}\}$. Let
\begin{displaymath}
S_{t}=\max_{\{i_{1},i_{2},\ldots,i_{t}\}\in \mathcal{P}_{t}(\{1,2,\ldots,m\})}
|V_{i_{1}}\cap
V_{i_{2}}\cap \cdots \cap V_{i_{t}}|.
\end{displaymath}
Combining the above expressions we have
$$\sum _{i=t}^{m}d_{i}{i \choose t}\leq S_{t}{{{m \choose t}}},$$
proving the theorem.
\end{proof}

\section{Concluding remarks}
\label{section:conclusion} When $m(p-1) < n \leq 2(m-1)(p-1)$, the
necessary and sufficient conditions for the existence of monochromatic
$K_{p,p}$'s in bipartite graphs remain open. It would be interesting to
see tighter results for $k$-partite graphs.

\section*{Acknowledgments}
We thank Gerard Jennhwa Chang for very helpful comments on a
preliminary version of this paper.

\bibliographystyle{abbrv}
\bibliography{bi}
\end{document}